\documentclass[prl,twocolumn,preprintnumbers,amsmath,amssymb]{revtex4-1}

\usepackage{graphicx}%
\usepackage{dcolumn}%
\usepackage{bm}%
\usepackage{amsmath, amsthm, amssymb} 
\usepackage{fontenc}
\usepackage{color,graphicx}

\begin{document}

\thispagestyle{plain}

\title{Critical Casimir forces and colloidal aggregation: A numerical study}

\author{Nicoletta Gnan}
\affiliation{Dipartimento di Fisica and CNR-ISC, Universit\`{a} di Roma ``La Sapienza'', 
Piazzale A. Moro $2$, $00185$ Roma, Italy}
\author{Emanuela Zaccarelli}
\affiliation{Dipartimento di Fisica and CNR-ISC, Universit\`{a} di Roma ``La Sapienza'', 
Piazzale A. Moro $2$, $00185$ Roma, Italy}
\author{Piero Tartaglia}
\affiliation{Dipartimento di Fisica and CNR-ISC, Universit\`{a} di Roma ``La Sapienza'', 
Piazzale A. Moro $2$, $00185$ Roma, Italy}
\author{Francesco Sciortino}
\affiliation{Dipartimento di Fisica and CNR-ISC, Universit\`{a} di Roma ``La Sapienza'', 
Piazzale A. Moro $2$, $00185$ Roma, Italy}

\vspace{0.6cm}

\begin{abstract}
We present a  numerical study of the effective potential $V_\mathrm{eff}$ between two
hard-sphere colloids dispersed in a solvent of interacting particles, for several
values of  temperature and solvent density,  approaching the solvent gas-liquid critical point. We investigate the stability  of a system of particles interacting via $V_\mathrm{eff}$ to evaluate the locus of colloidal aggregation in the solvent phase-diagram, and its dependence on the colloid size. 
We assess how the excluded volume depletion forces are modified by solvent attraction and 
discuss under which conditions  solvent critical fluctuations, in the form of critical Casimir forces, 
can be used to effectively manipulate  colloidal aggregation.  
\end{abstract}

\maketitle

\vspace{0.5cm}


\footnotetext{$\ast$~\textit{E-mail:nicoletta.gnan@roma1.infn.it}}
\footnotetext{\textit{$^{a}$~Dipartimento di Fisica, Universit\`{a} di Roma ``La Sapienza'', 
Piazzale A. Moro $2$, $00185$ Roma, Italy }}
\footnotetext{\textit{$^{b}$~Dipartimento di Fisica and CNR-ISC, Universit\`{a} di Roma ``La Sapienza'', 
Piazzale A. Moro $2$, $00185$ Roma, Italy }}



\section{Introduction}
Understanding effective interactions  between colloidal particles in the presence of macromolecular additives (e.g. surfactants, polymers, micelles, etc.) is of fundamental importance in  soft-matter science.  Since the pioneering works of Asakura and Oosawa (AO)~\cite{AO} and Vrij~\cite{Vrij}, who modeled the additives as  an ideal gas depletant, it is known that the presence of (small) co-solutes in solution gives rise to entropic forces that significantly modify the structure and the dynamics of colloidal systems ~\cite{Mao,Roth}. These depletion forces arise from the entropy-driven exclusion of the co-solute from the volume between particles when their  surface-to-surface relative distance is comparable to the co-solute size~\cite{Lykos}.  Accurate numerical~\cite{Dickman, Dijkstra} and theoretical ~\cite{Attard1,Attard2, Gotzelmann} calculations of the effective potential between  hard-sphere colloids dispersed  in a depletant of small hard-sphere (HS) particles 
have shown that the entropic attraction arising at short distances
is  followed by an oscillatory behavior due to excluded volume correlation effects. 
 
It is also well known that depletion forces can be strong enough to induce colloid aggregation. Decreasing the size ratio $q$ between the hard-sphere depletant (of diameter $\sigma_s$) and the hard-sphere colloid   ($\sigma_c$)  gives rise to a 
critical point~\cite{Dijkstra} at which colloids separate in a colloid-rich and a colloid-poor phase, the analog of the gas-liquid separation in one-component systems.   Interestingly, the smaller  $q \equiv \sigma_s/\sigma_c$, the smaller is the volume fraction of small particles  $\phi_s=(\pi/6)(N_s/V)\sigma_s^3$ (where $N_s$ is the number of depletant particles and $V$ is the available volume they can access) required to induce phase separation, bringing to the
paradoxical predictions that it would seem impossible to dissolve colloids in the presence of 
extremely small additives.
Clearly, although modeling all interactions
in terms of excluded volume provides a model amenable to theoretical and numerical treatment, this approach does not fully capture the ingredients behind effective potentials in real systems. This motivates recent attempts to include colloid-depletant~\cite{foffijpcb114} and depletant-depletant interactions~\cite{egorovjcp115, jamnik1,amokranepre66, jamnik2}. 
It has been shown that the latter  significantly affect the resulting effective potential, altering the strength of the attraction and, in some cases,  
suppressing the oscillatory behavior characteristic of the
depletant granularity~\cite{egorovjcp115, jamnik1,amokranepre66, jamnik2}.  

A more realistic description of depletant-depletant interactions brings in two important features:
first,  beside  $\phi_s$, the temperature $T$ becomes a relevant variable for modulating the effective potential;  second, the depletant itself can undergo a gas-liquid phase separation. Close to the critical point of the depletant, critical fluctuations  $-$with the associated critical Casimir forces~\cite{FisherDeGennes, gambassipre80, hertleinnat451}$-$ might become the dominant contribution in the effective potential --with respect to the depletion force -- hence driving  colloidal aggregation~\cite{beysensjsp95, Guoprl100, bonnprl103}.
For this to take place, as recently pointed out~\cite{gambassiprl105}, it is necessary that the colloidal solution remains stable outside the depletant   critical region, i.e. that Casimir-driven phase-separation is not preempted by other effects.
A recent experimental 
work~\cite{Buzzaccaro} has addressed the role of critical fluctuations in colloidal aggregation.
In this study colloidal particles were dissolved in a micellar solution which is known to undergo a gas-liquid phase separation upon heating.  Exploring the onset of colloidal aggregation as a function of $T$ and micellar concentration, the authors were able to show that a continuous line can be drawn in the $(\phi_s,T)$ plane separating the region where  colloids are stable from the region where colloids aggregate. This line was found to monotonically connect the  pure depletion
limit  to the micellar critical point, suggesting the possibility that critical fluctuations  can be considered as an extreme case of depletion forces where the dense critical regions act as renormalized depletants, thus proposing a link between depletion forces and critical Casimir effect~\cite{Buzzaccaro}.
Although this interpretation in terms of  a unique description of forces originated by different mechanisms is fascinating, critical Casimir forces have a richer phenomenology when compared to standard depletion forces. Indeed critical Casimir forces depend strongly on 
the interaction between the colloid and the critical depletant  and it is also possible to generate repulsive Casimir forces\cite{hertleinnat451}, a case which does not have a depletion analog. 

Numerical simulations of simple models can help understanding if, how and when
critical fluctuations may play a role in colloidal aggregation.
 To this aim, in this work we evaluate numerically the effective potential between
two hard-sphere colloids immersed in an implicit solvent, in the presence of interacting depletant  particles,  for a wide region of state points $(\phi_s,T)$, including the critical region. We investigate  two different depletant models at several values of $q$, between $0.05<q<0.2$,   at the limit of todays computational possibilities,  in order to disentangle aggregation driven by critical
 fluctuations from that arising from different phenomena. We provide 
a complete evaluation of the colloid stable and unstable regions 
and the relative location of these regions with respect to the depletant gas-liquid critical point as a function of $q$. Our main result is that critical fluctuations  
are responsible for colloidal aggregation only under very specific conditions. 

\section{Models and Methods}  
We calculate the effective forces between two hard-spheres immersed in two different models for the depletant  particles. In the first model, depletant particles interact via pairwise square-well potential (SW)
\begin{equation}
\Phi^{SW}_{ij}(r_{ij})=\begin{cases} \infty, & r<\sigma_s \\ 
-\varepsilon, & \sigma_s\leq r< (1+\Delta) \sigma_s \\ 
 0, & r\geq (1+\Delta)\sigma_s\\
\end{cases}
\end{equation}
\noindent  with a well width  $\Delta=0.1$ and a well depth $\varepsilon$. 
The second depletant model is an anisotropic three-patches (3P) Kern-Frenkel~\cite{Kern}system which consist of hard-sphere particles decorated with three attractive sites whose interaction potential can be defined as:
\begin{equation}
\Phi^{3P}_{ij}(\mathbf{r}_{ij};\hat{n}_i,\hat{n}_j)=\Phi^{SW}_{ij}(r_{ij})\cdot f(\hat{r}_{ij};\hat{n}_i,\hat{n}_j)
\end{equation}
\noindent i.e. as a SW potential modulated by an angular function
\begin{equation}\label{eq:angular_func}
f(\hat{r}_{ij}; \hat{n}_i,\hat{n}_j)=\begin{cases} 
1 , & if
\begin{cases} &\hat{r}_{ij}\cdot \hat{n}^{(\alpha)}_{i} \geq \cos(\theta)\\
and & -\hat{r}_{ij}\cdot \hat{n}^{(\alpha)}_{j} \geq \cos(\theta)
\end{cases}\\
0 & otherwise .
\end{cases}
\end{equation}
\noindent In Eq.\ref{eq:angular_func}  $\hat{n}^{\alpha}_{i(j)}$ is the unit  vector associated to  the orientation of  patch $\alpha=1,2,3$ on particle $i(j)$ and $\hat{r}_{ij} \equiv {\vec r_{ij}}/{|\vec r_{ij}|}$, where 
$\vec r_{ij}$  is the vector connecting the centers of two particles. The value of $\cos \theta=0.8947$ controls the width of the patch, while the value of $\Delta=0.119$ fixes the interaction range. Therefore two patchy particles interact via an attractive SW potential  only if the two patches are properly aligned. 
The depletant particle diameter $\sigma_s$ and the potential well $\epsilon$ (equal for both models) are chosen as unit of length and energy.  $T$ is measured in units of  $\epsilon$  (i.e. Boltzmann constant $k_B=1$). Both models reduce to the HS model at high $T$, when attraction becomes irrelevant.  The SW and the 3P models
are characterized by a gas-liquid critical point,  located respectively at ($T_c = 0.478, \phi_c = 0.25$~\cite{largojcp128}) and ($T_c = 0.1247, \phi_c = 0.073$).  
The difference in the critical $\phi$ between the two models, due to the very small $\phi_c$ of the 3P model, provides a different behavior for the solubility of colloids along the critical isochore of the two depletants.
The investigation of both models thus offers the possibility to establish this behavior  for the case of a simple depletant (SW) as well as for a depletant characterized by a  small value of $\phi_c$ as in the experimental system of Ref. 23.
We  perform  Monte Carlo simulations of two
colloids fixed at several equally-spaced distances in a depletant of
SW or 3P particles, as a function of $T$ and  depletant (reservoir) concentration. 
The size of the rectangular box has been chosen in such a way that the depletant density far from the colloids reaches the constant reservoir (bulk) value, in all directions. The effective force $F_\mathrm{eff}$ is evaluated by performing virtual displacements of  $\Delta r$ (both positive and negative) for each colloid  and  computing the probability of having at least one collision $P_\mathrm{overlap}(\Delta r) \equiv  \langle 1-e^{-\beta \Phi_{hs}}\rangle $ with the depletant particles.  As shown in Ref. ~\cite{Bratko}: 
  
 \begin{equation}
 \beta F_\mathrm{eff}(r) = - \lim_{\Delta r \to 0^+} \frac{  P_\mathrm{overlap}(\Delta r) }{\Delta r} - \lim_{\Delta r \to 0^-} \frac{ P_\mathrm{overlap}(\Delta r) }{\Delta r}
 \end{equation}

\noindent We have chosen $\Delta r$ in such a way that $P_\mathrm{overlap}(\Delta r) \approx 5\%$. The effective colloid-colloid potential  $\beta V_\mathrm{eff}(r)$  is then obtained by integrating  $\beta F_\mathrm{eff}(r)$ from $\infty$ to $r$.  
In addition, we perform standard MC simulations of the bulk SW model to numerically study the critical region. Also, to assess colloidal stability upon varying depletant conditions, 
we perform grand canonical Monte Carlo simulations of colloidal particles interacting via $V_{eff}$ for 
several values of $q$ and $\phi_c$.

\section{Results} 
\subsection{Critical behavior of the depletant}
Before discussing the results for the effective forces, we investigate the critical region of 
the pure depletant solution.  To quantify the  growth of the thermal correlation length $\xi$ on approaching the critical point we perform  simulations of a bulk SW model along the critical isochore\cite{largojcp128} for several $T$ 
values and evaluate numerically the spherically averaged structure factor $S(k)$~\cite{Hansennew}, shown in Fig.~\ref{fig:S(k)}. The correlation length $\xi$ is estimated fitting the low wave vector $k$ behavior of 
$S(k)$ according to
\begin{equation}
 S(k)=c+\frac{S(0)-c}{1+\xi^2 k^2}.
 \label{eq:Sq}
\end{equation}
In Eq.~\ref{eq:Sq}, a constant $c$ is added to the critical Lorentzian shape to account for the 
low $q$ non-critical component, which is still not negligible at the highest investigated $T$. 
The best fitting functions are reported also in Fig.~\ref{fig:S(k)}.

\begin{figure}[h]\label{fig:S(k)}
\centerline{\includegraphics[width=.9\linewidth]{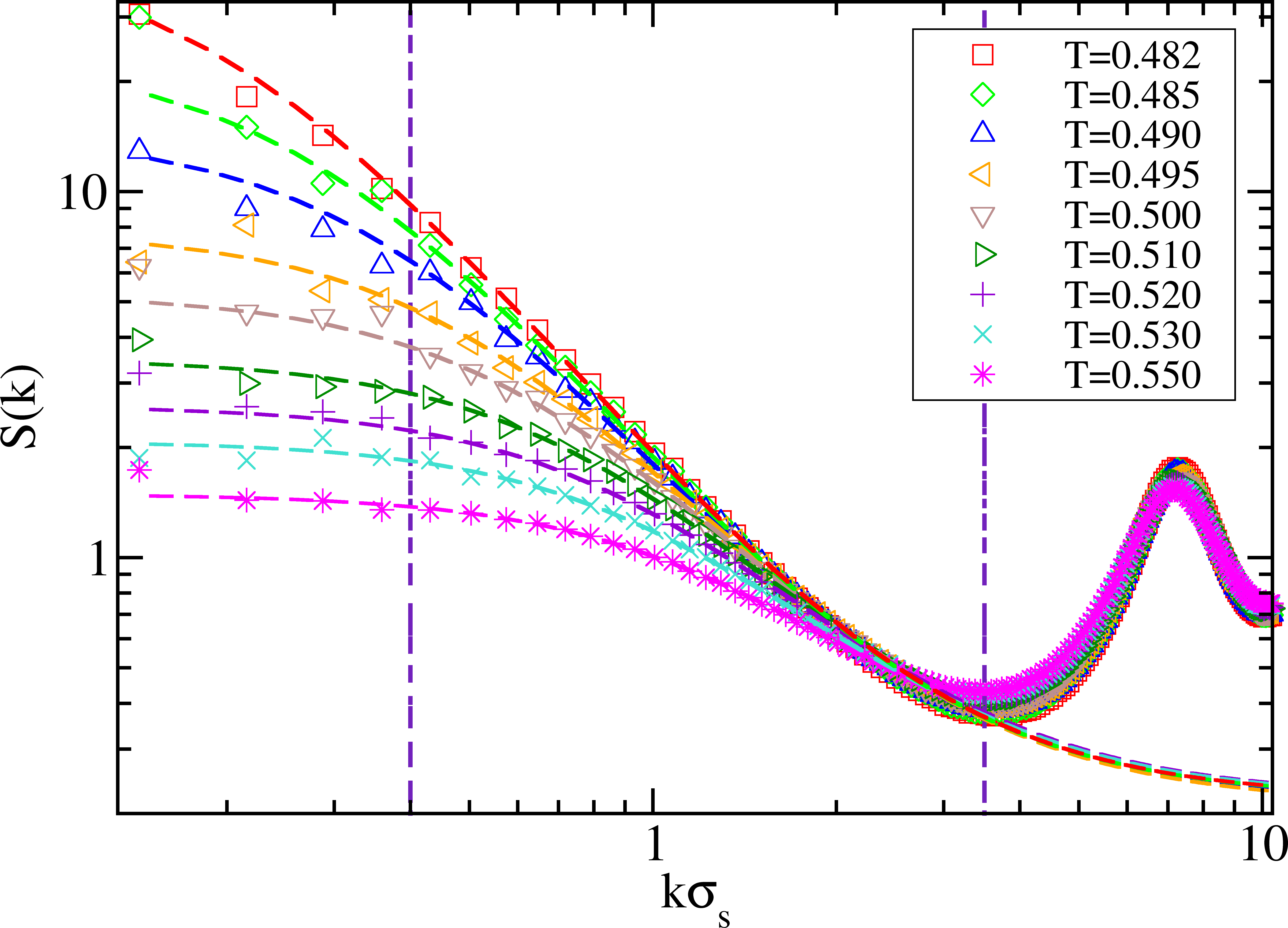}}
\caption{\label{fig:S(k)} Static structure factors $S(k)$ (symbols) for a bulk SW system of $40000$ particles along the critical isochore as a function of $T$.  Dashed lines are fit according to 
 Eq.\ref{eq:Sq}.  The fitting parameter $c$ of Eq.\ref{eq:Sq} is found to be constant at all $T$ and equal to $c\simeq 0.22$. The fit is restricted to a fixed interval (for all $T$) delimited by the two vertical dashed lines ($0.4\leq k\sigma_s\leq 3.5$). The small $k$ values of $S(k)$ are indeed affected by numerical errors related to the small number of independent wave vectors contributing to the spherical average.  The large $k$ cut-off  excludes the non-critical part of $S(k)$. }
\end{figure}

\noindent In the critical region a universal power-law behavior is expected for
$\xi$ and $S(0)=\rho k_B T \chi_T$ (where $\chi_T$ is the compressibility)~\cite{Hansennew}:
\begin{eqnarray}
\label{eq:exponents}
\xi & \simeq &\xi_0   \left( \frac{T-T_c}{T_c}   \right)^{-\nu}\\
S(0) & \simeq & C_0\left( \frac{T-T_c}{T_c}   \right)^{-\gamma}.
\end{eqnarray}
with $\nu=0.63$ and $\gamma=1.24$, since the SW model belongs to the  Ising universality class. $\xi_0$ and $C_0$  are the non-universal critical amplitudes.

\begin{figure}[h]
\centerline{\includegraphics[width=.9\linewidth]{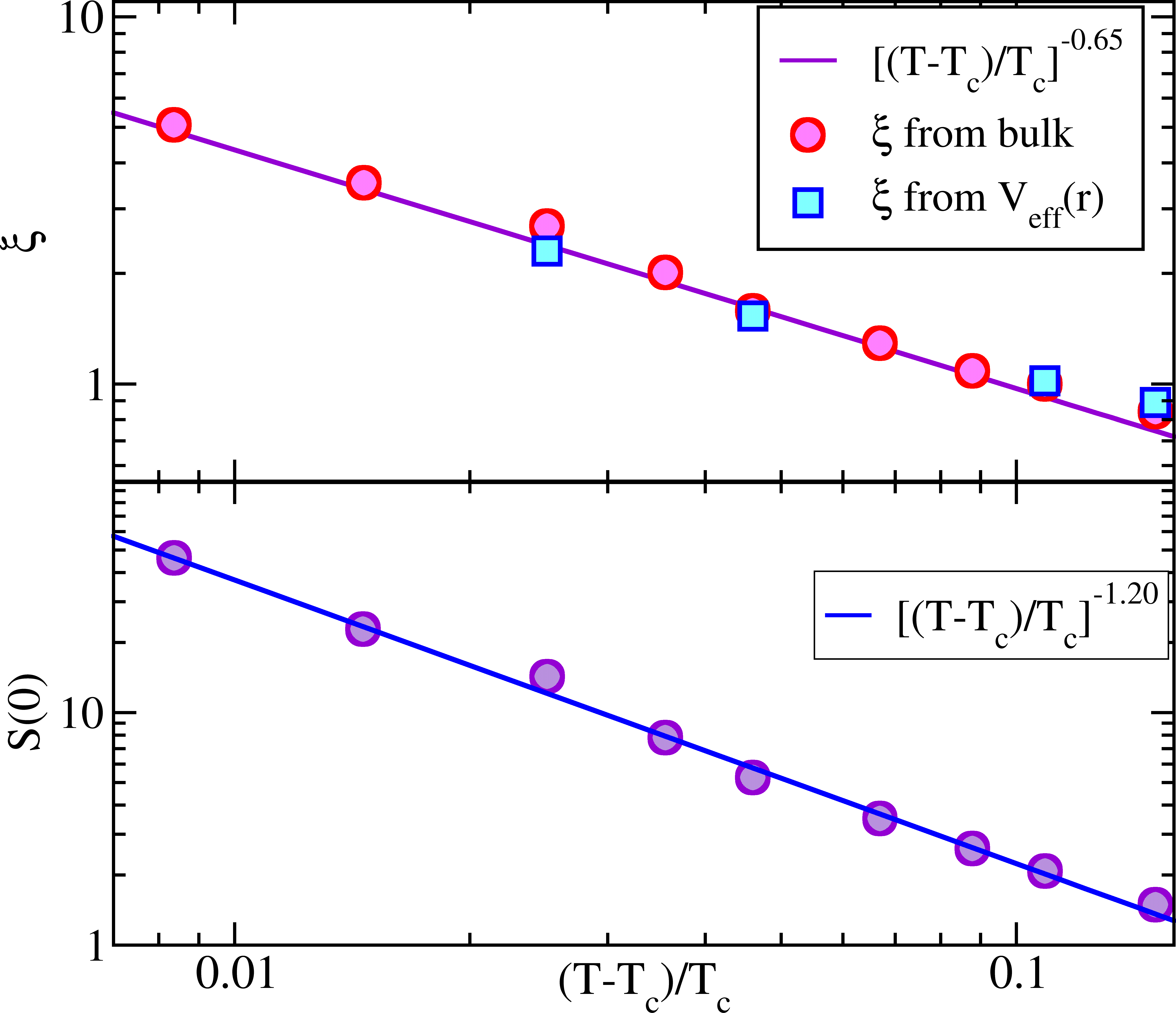}}
\caption{Power-law behavior of the thermal correlation length (top panel) and $S(0)$ (lower panel) approaching the critical point for a  bulk SW system (circles), resulting from fitting $S(k)$ according to
Eq.~\ref{eq:Sq}.
Lines are power-law fits, with best fit exponents $\nu=0.65$ and $\gamma=1.20$, in good agreement with the expected Ising values. The best fit value for the amplitude $\xi_0 \sim 0.23 \sigma_s$.
Squares indicates the characteristic length scale  obtained from fitting the effective potential according to Eq.~\ref{eq:Casimir}.
\label{fig:scaling} 
}
\end{figure}

The two panels in Fig.~\ref{fig:scaling} show that both $\xi$ and $S(0)$ indeed satisfy  Eq.~\ref{eq:exponents} with 
$\nu$ and $\gamma$ values  consistent with   the expected Ising ones, when 
 $(T-T_c)/T_c\lesssim 0.1$.  Thus, the onset of the critical region can be defined as $T \approx 1.1 T_c$.

\subsection{Effective Potentials}
\begin{figure}[h]
\centerline{\includegraphics[width=.9\linewidth]{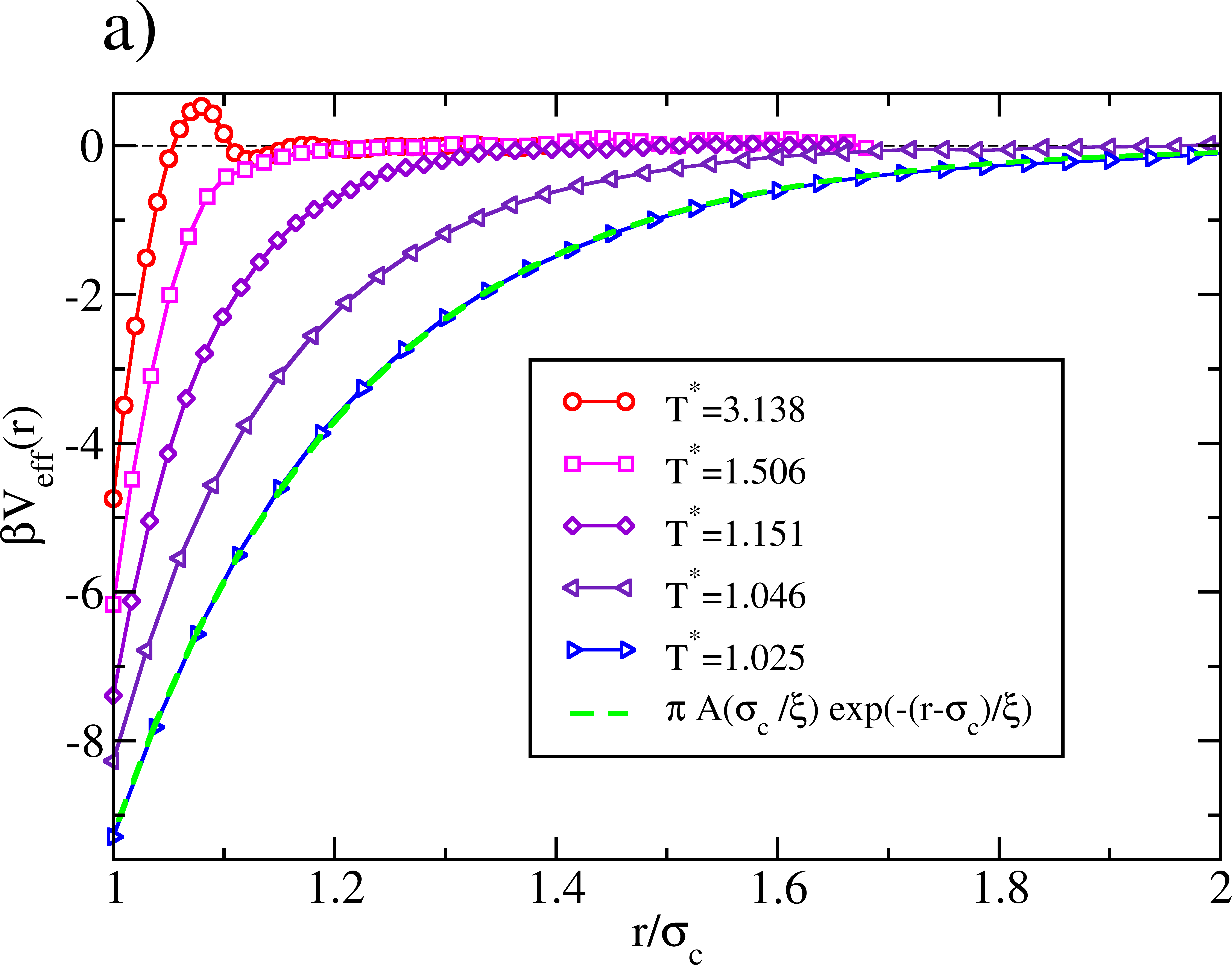}}
\centerline{\includegraphics[width=.93\linewidth]{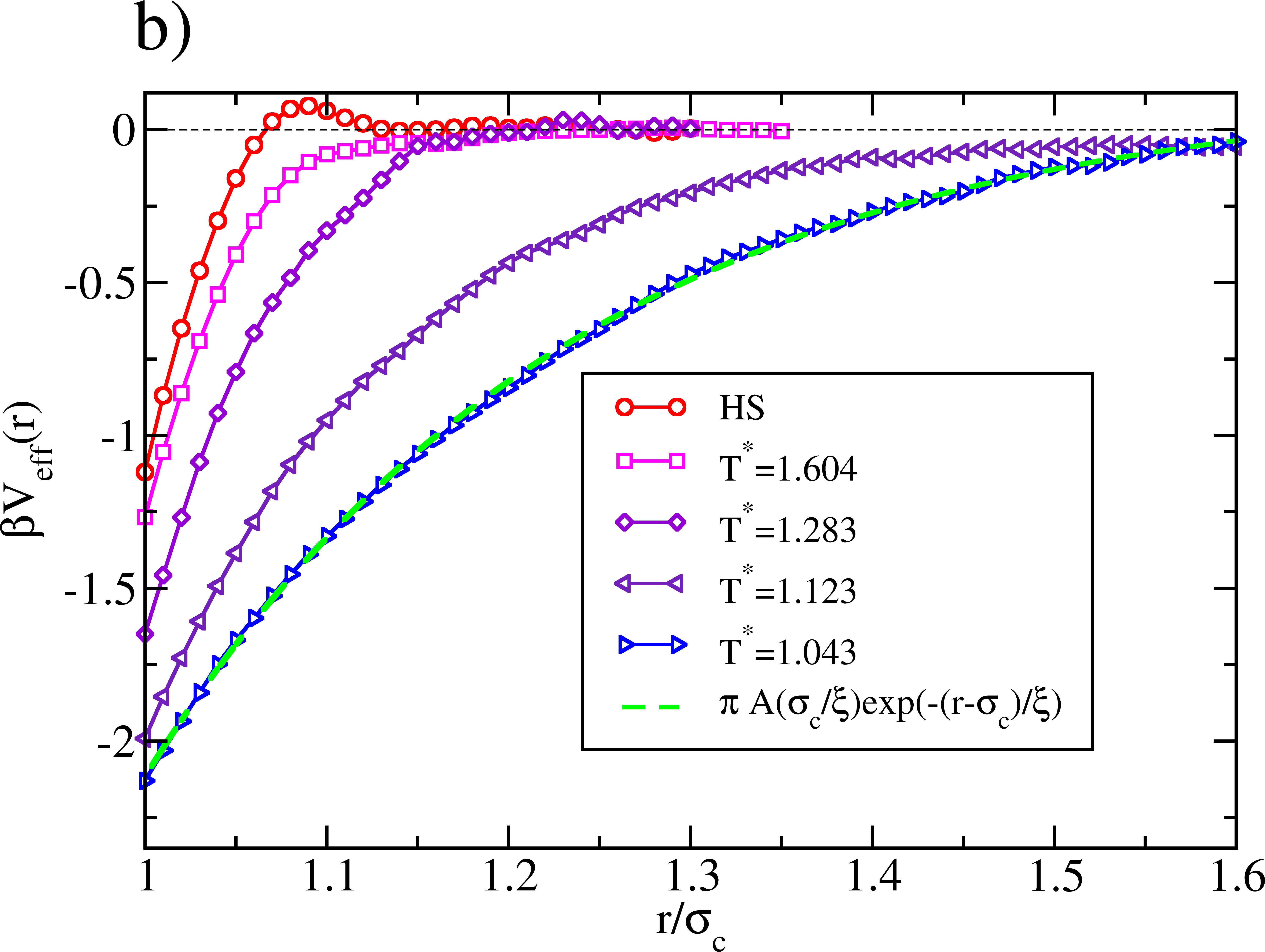}}

\caption{\label{fig3}
Effective potential for  $q=0.1$ for the (a) SW and (b) $3$P potentials along their respective
critical isochores for different values of $T^*= T/T_c$. Dashed lines for $T^*<1.045$ are
fits according to Eq.~\ref{eq:Casimir},  with fit parameters $A_{SW}=-0.63$, $A_{3P}=-0.19$ and $\xi_{SW}=2.23$, $\xi_{3P}=2.28$. }
\end{figure}

Fig.~\ref{fig3}  shows the evolution of the effective potential along the reservoir critical isochore. 
As previously observed for the Baxter model~\cite{jamnik1}, $V_\mathrm{eff}$  
progressively loses its oscillatory character  upon cooling and gradually turns into a completely attractive 
potential.  In addition, a progressive significant increase of the interaction range is observed on approaching $T_c$, signaling the onset of critical Casimir forces.  A continuous path thus connects the high-$T$ pure HS depletion to the Casimir interaction.
Casimir forces are very dependent on the geometry in which the critical fluid is confined as well on the bulk universality class of the (confined) fluid and the boundary conditions of the confining surfaces~\cite{gambassipre80,Vasilyev}.
Recently it was shown that in the case of a critical binary mixture confined between two spherical surfaces of diameter $\sigma$ at  minimum distance between the two surfaces $z \equiv r-\sigma$, the potential generated by the critical Casimir forces can be written as
\begin{equation}\label{eq:Phi_z}
\beta \Phi(z)=\frac{\sigma}{z}\Theta(z/\xi).
\end{equation}
Eq.\ref{eq:Phi_z} is valid only if $\sigma \gg z$ , i.e. if one can exploit the Derjaguin approximation ~\cite{derjaguin}. $\Theta(z/\xi)$ is a scaling function which contains all the information about the critical properties of the system, the geometry and the boundary conditions, while  $\xi$ is the thermal correlation length. The sign of the Critical Casimir forces  (repulsive or attractive) depend on the 
so called boundary conditions, i.e. on the  ability of the two large colloids of absorbing or not the depletant.
In the present case, the interaction between the colloid and the SW or 3P particles is
purely repulsive (hard-sphere). We indicate in the following such condition with the symbol $(--)$. 
The long distance behavior of  (sphere-sphere) $\Theta(z/\xi)_{(--)}$ (i.e. for $z/\xi \gg 1$) is~\cite{gambassipre80} 
\begin{equation}\label{eq:Theta_minus}
\Theta(z/\xi)_{(--)} (z/\xi\gg 1) =  \pi A (z/\xi) e^{-(z/\xi)}
\end{equation}
\noindent where A is a universal constant.
It follows that the long distance behavior of the
critical Casimir potential between the two spherical colloids  behaves as
\begin{equation}
\beta \Phi(r)  = \pi A \frac{\sigma_c}{\xi} e^{-(r-\sigma_c)/\xi}.
\label{eq:Casimir}
\end{equation}

Fig. \ref{fig3} shows that close to criticality,  the effective potential $\beta V_{eff}$ for both the SW and 3P models can be represented  by  Eq.\ref{eq:Casimir}, in the entire range of distances.
Fitting the  low $T$ numerical results for $\beta V_{eff}$ with Eq.\ref{eq:Casimir} it is possible to extract a value of $\xi$ that should correspond to the correlation length of the critical bulk fluid.

\noindent To show that the critical Casimir forces are controlling the effective interactions near $T_c$ we compare in Fig.~\ref{fig:scaling} the $\xi$ value resulting from the fit of  the effective potentials with that of the bulk SW system. 
The good agreement between the two values of $\xi$ strongly supports the idea that 
the effective potential close to the critical point is controlled by critical fluctuations.
\begin{figure}[h]
\centerline{\includegraphics[width=.9\linewidth]{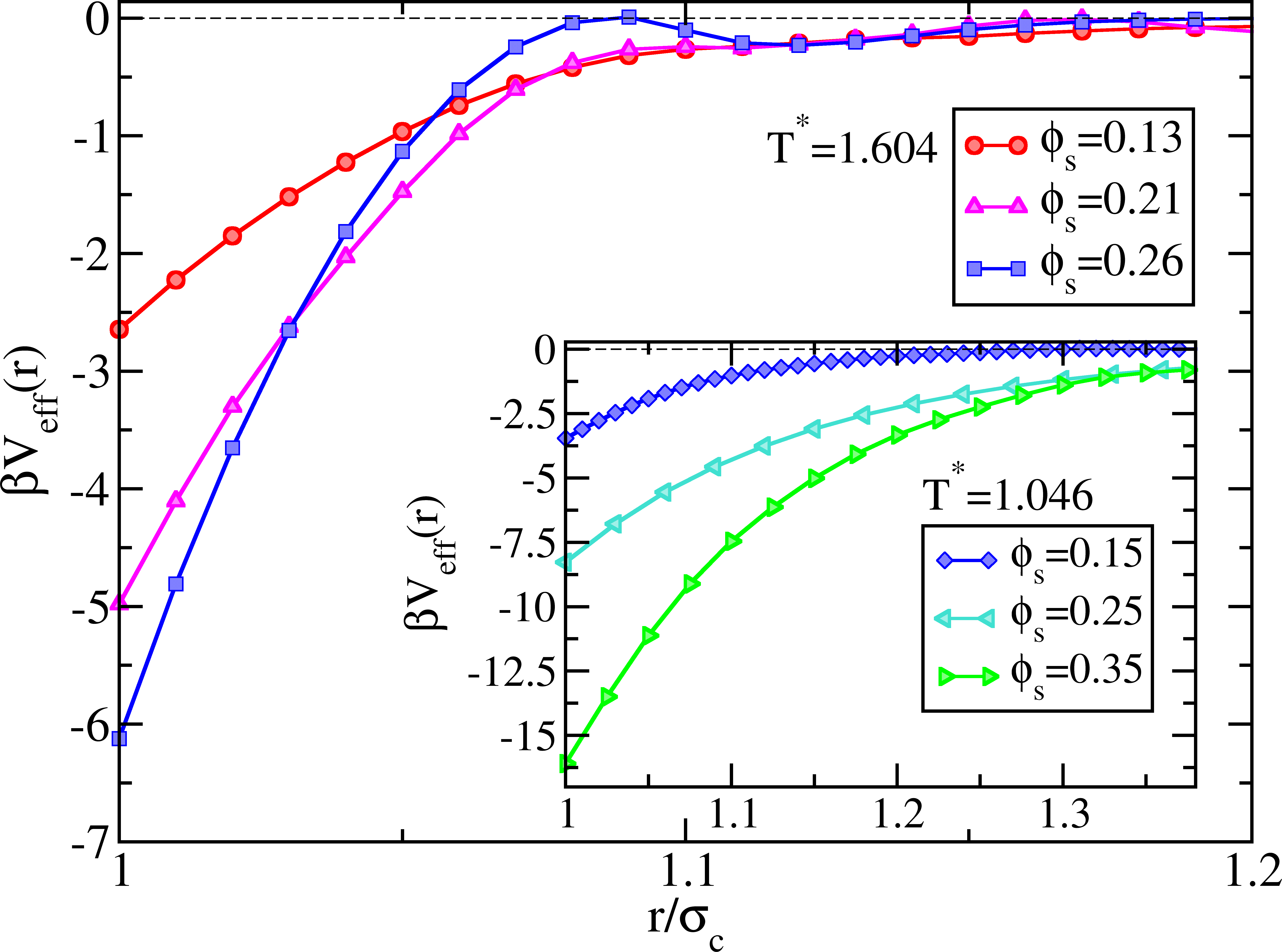}}
\caption{\label{fig:pot_dens}
Effective potential for $q=0.1$ for the $3$P potential and the SW (inset) for three different values of
the reservoir  $\phi_s$, along an isotherm.}
\end{figure}

\noindent For completeness, Fig. \ref{fig:pot_dens} shows the dependence of $V_{eff}$ on depletant concentration for the 3P model when $q = 0.1$. The absolute  value of the effective potential at contact becomes larger with increasing $\phi_s$, due to the larger amount of depletant in suspension. Instead, the interaction range has a maximum at the critical isochore, significantly increasing only on approaching $T_c$, retaining the typical depletion interaction range when far from the critical point. The inset of Fig. \ref{fig:pot_dens} shows similar data for the SW model.
 
\subsection{Phase separation of colloids: role of the depletant size}
To evaluate the stability of the colloidal solution in different regions of the 
depletant (reservoir)  phase-diagram we investigate a system of colloidal particles
interacting via  the pair-wise additive Hamiltonian $H=\sum_{ij} V_{HS}(r_{ij})+V_\mathrm{eff}(r_{ij})$ (where $r_{ij}$ is the relative distance between two generic particles $i$ and $j$). While the assumption of pairwise additivity is essentially uncontrolled, 
since the presence of a third particle in the vicinity of a pair of colloids  will alter the depletant spatial distribution,  it is expected that such approximation becomes progressively less relevant on decreasing $q$ and that its main effect is to slightly shift the coexistence line  towards higher values of  the depletant density in the reservoir ~\cite{amokranemolphys99}.
We thus perform grand canonical MC simulations (GCMC)   for several values of the 
chemical potential, spanning the entire density region.  In this way we are able to 
detect if particles interacting via $\beta V_{HS}+\beta V_\mathrm{eff}$  form a stable  fluid phase 
for all densities 
or if there exist a density region where particles  phase separate in  colloidal-rich colloidal-poor coexisting phases, the analog of the gas-liquid coexistence.  Repeating such calculations for  the effective potential evaluated at different  $\phi_s$ and $T$ state points, 
the depletant bulk phase diagram can be separated into a  region in which colloids are 
in a homogeneous stable fluid phase for all densities  from a region in which colloids undergo phase separation.

\begin{figure}
\centerline{\includegraphics[width=.9\linewidth]{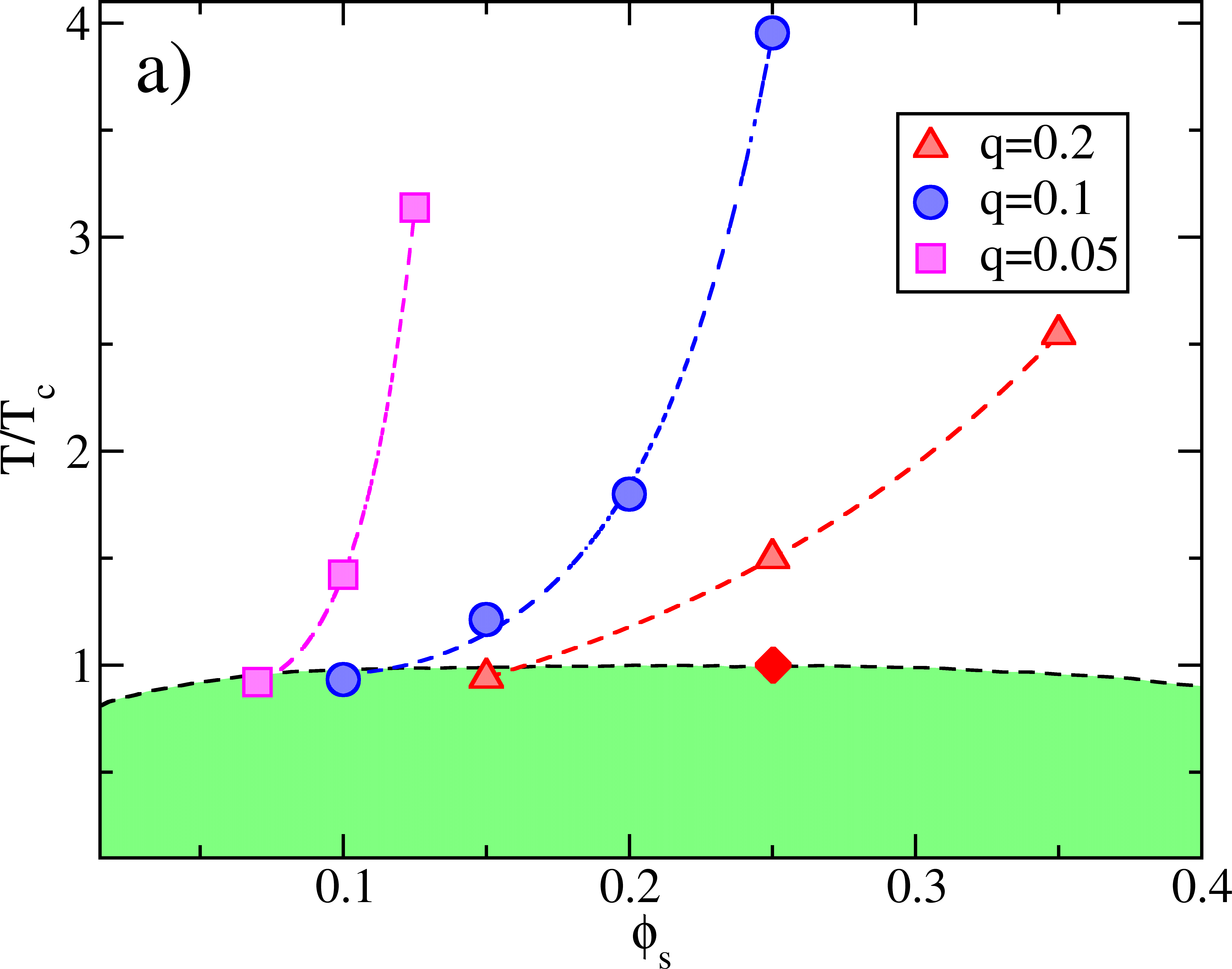}}
\vspace{1.5cm}
\centerline{\includegraphics[width=.9\linewidth]{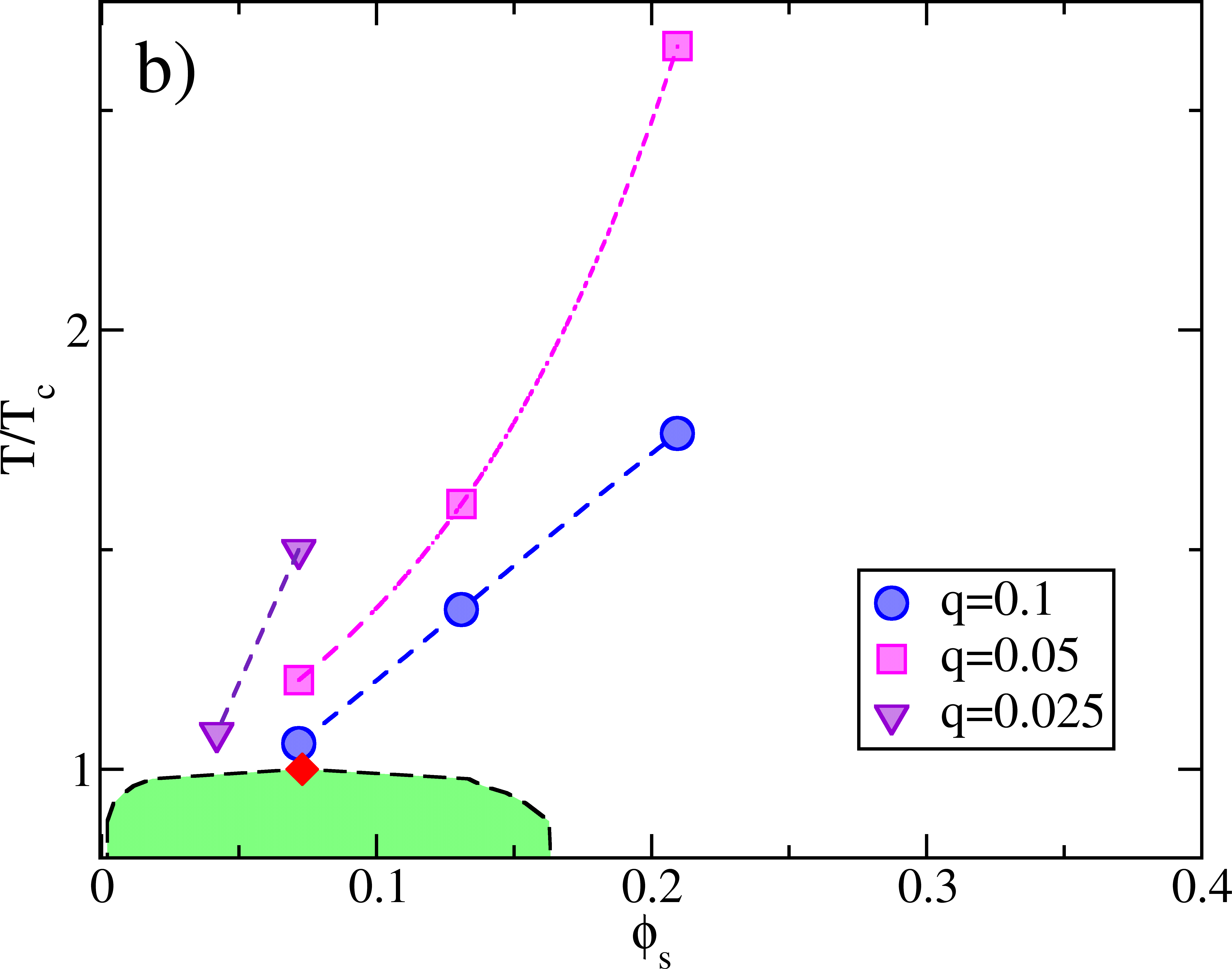}}
\caption{\label{fig:phaseSW}
Phase diagram of (a) the SW model and of (b) the 3P model in the $T/T_c$ and reservoir packing fraction $\phi_s$ plane. The diamond indicates the critical point. The area below the critical point (green) is the phase separation region. Curves interpolating circles, squares and triangles are the loci of colloidal aggregation for several $q$ values.  The $q=0.025$ case in (b) has been calculated assuming the validity of the Derjaguin approximation. The dashed lines connecting the points are guides to the eye.
The reduced temperature at which the solvent second virial coefficient is zero is $T^{*}=1.506$ for the SW model and $T^{*}=1.738$ for the 3P model.
}
\end{figure}
The loci separating fluid from phase separated state points, according to GCMC calculations, for different $q$ values are displayed in Fig.~\ref{fig:phaseSW}  in the phase diagrams of both depletant models. In each figure the  gas-liquid critical point and the phase coexistence region  are reported.  
Each locus starts from the HS pure depletion limit and converges toward the critical point, ending on the gas side of the depletant gas-liquid coexistence curve.
In the case of the SW, for $q=0.1$, the crossover $T$ at which the solution starts to be unstable along the critical isochore is much higher than the $T$ of the onset of critical regime. Colloidal phase separation in this case is therefore not driven by depletant criticality but by standard depletion forces.
In the case of the $3P$ depletant of Fig.~\ref{fig:phaseSW} (b), the aggregation line for $q=0.1$ intersect the gas-liquid coexistence  very close to the depletant critical point, suggesting   that, at the depletant critical density, it is possible to go close enough in $T$ to the critical point to induce colloid phase separation via Casimir forces. We remark that such a difference originates from the different critical packing fraction between the two models.

We also study how the locus of onset of instability depends on  $q$. 
To this aim, we repeat our calculations for $q=0.05$ (for SW also $q=0.2$), and the
corresponding results are also shown in Fig.~\ref{fig:phaseSW}.
At this small $q$ value, more than $30.000$ depletant particles are needed in the simulations, 
requiring a significant numerical investment.  It is possible to extend these calculations to smaller $q$ values (where simulations becomes prohibitive due to the large number of solutes) by invoking  the  Derjaguin approximation~\cite{derjaguin}, according to which~\cite{Lekkerbook} for two different size ratios $q_1$ and $q_2$ holds: 
\begin{equation} \label{eq:derj}
\frac{ V_\mathrm{eff}(h/\sigma_s,q_1)}{\sigma_1}= \frac{V_\mathrm{eff}(h/\sigma_s,q_2)}{\sigma_2} 
\end{equation}
where $h$ is the surface to surface distance. Eq. \ref{eq:derj} is expected to hold accurately as far as the curvature radius of the interacting spheres is  sufficiently larger than the range of the interaction potential.

 We find (not shown) that when the ratio between the interaction range and the colloid diameter is smaller than 10\% the approximation is rather good. In principle, we can thus   extend the $q=0.05$ data to smaller $q$ and draw stability lines  for arbitrary values of $q<0.05$, as shown for example in Fig.~\ref{fig:phaseSW}(b) for $q=0.025$. Our results show that  $q$ plays a significant role in the location of the instability line, in agreement with the 
 results reported in a pioneering, but largely unnoticed, experimental study~\cite{piazzapps100}. 

Finally we notice that along the stability lines the value of the reduced second virial coefficient, i.e. the second virial coefficient normalized to the hard-sphere value~\cite{Noro_00},
\begin{equation}\label{eq:B2}
B_2^{*} = \frac{3}{\sigma^3_{c}} \int_{0}^{\infty} [1-\exp(-\beta V_{eff})]r^2 dr
\end{equation} \noindent 
is close to the value $B_2^{*} \approx -1.6$ which has been identified as the universal
value for the onset of gas-liquid separation in short-range attractive potentials~\cite{Noro_00}.

\section{Discussion and Conclusions}  Several considerations can be drawn from the analysis of the
data shown in the previous figures.  ($i$) The depletant-depletant attraction always
favors colloidal aggregation as shown by the monotonic behavior of the curves in Fig.~\ref{fig:phaseSW}.
Along the critical isochore, approaching the critical $T$, $\beta V_\mathrm{eff}$
becomes increasingly long-ranged and its radial dependence is well represented by an exponential decay.  This decay length  is controlled by $\xi$, in agreement with  theoretical predictions (Eq.~\ref{eq:Casimir})\cite{gambassipre80}.  The amplitude  
is found to be $T$-dependent  stressing the importance of the condition $\xi \gg \sigma_s$ for fully entering in the theoretically expected scaling regime.  
($ii$) $\beta V_\mathrm{eff}$  has been used to evaluate the
colloidal aggregation loci under the pairwise additivity approximation. We find that the $q$-dependence of these lines is quite strong. 
Indeed, each locus starts from the high $T$ HS limit, 
which provides a lower value of $\phi_s$ for smaller $q$ ~\cite{Dijkstra}, and bends towards lower $\phi_s$ until merging with the depletant gas-liquid coexistence, in agreement with  the experimental work of Ref. \cite{Buzzaccaro}.
 The marked dependence of the locus of aggregation on the size of the colloid strongly suggests that no universal connection can be exploited between specific thermodynamic properties of the depletant (which are $\sigma_c$ independent) and aggregation. 
Interestingly, for the SW depletant,  colloids as small as ten times the depletant size (see $q=0.1$ or smaller in Fig.~\ref{fig:phaseSW}a), aggregate well before the depletant develops significant critical fluctuations, proving that colloidal aggregation driven by
a critical depletant requires specific conditions to be observed~\cite{gambassiprl105}. However,  depletants with a low $\phi_c$ (as exhibited by particles with limited valence interactions~\cite{bian,Zacca1}) enhance the possibility of observing Casimir aggregation 
since the underlaying  depletion interaction, whose amplitude is controlled by $\phi_s$, is  weaker. This is indeed the case for the 3P model for $q=0.1$ (see Fig.\ref{fig:phaseSW}b) as well as for the experimental system investigated in Ref. 23. ($iii$) Finally, we observe that critical Casimir forces may play a dominant role in colloidal 
aggregation  when a colloid-colloid repulsive interaction compensates the
non-critical component of the depletant mediated attraction (which is limited to
the $\sigma_s$ scale).   Indeed,  the effective potential calculated here can also  be 
used  to evaluate the stability line when the colloid-colloid hard-sphere interaction is
complemented by additional terms. When an additional repulsive potential of the order of a few $k_BT$, acting on the length scale of a few depletant diameters, is present (the typical case for screened electrostatic interactions), we find that colloidal aggregation takes place {\it only}  very close to the depletant critical point (i.e. only when $\xi \gg \sigma_s$), where the  critical Casimir potential, which decays on the scale of $\xi$, becomes the leading term in $\beta V_\mathrm{eff}$.  
Hence, our study goes beyond the purpose of using colloids as model systems for studying Casimir forces ~\cite{hertleinnat451,soykaprl101} since it suggests that with an appropriate tuning of electrostatic interactions and depletant conditions, critical Casimir forces can be exploited to finely control collective assembly\cite{bonnprl103,Buzzaccaro}.
The sensitivity of the  depletant critical fluctuations, where strength and range of interaction can be significantly changed via a tiny $T$ variation, opens up  a new way of manipulating colloidal aggregates which need to be exploited in the near future.


         

 {\it Acknowledgments:} We acknowledge support from MIUR-PRIN, ERC-226207-PATCHYCOLLOIDS and ITN-234810-COMPLOIDS.  We thank A. Gambassi,  A. Parola, 
 R. Piazza and S. Buzzaccaro for  interesting discussions.
 \bibliography{biblio_patchy}
\bibliographystyle{src}
\end{document}